\documentstyle[12pt]{article}
\input epsf.sty
\title{Order-chaos transitions in field theories with topological terms: a dynam
ical systems approach}

\author{{\bf C.Mukku}\\Dept. of Mathematics and Statistics\\
School of Mathematics and Computer/Information Sciences\\
University of Hyderabad,Hyderabad-500 046, India.\\
{\bf M.S.Sriram and J. Segar}\\Dept. of Theoretical Physics, 
University of Madras \\ Madras-600 025, India,\\
{\bf Bindu A. Bambah}\\ School of Physics\\ University of Hyderabad, 
Hyderabad-500 046, India\\
{\bf S. Lakshmibala}\\ Dept. of Physics, Indian Institute of Technology, 
\\ Madras-600 036, India}

\date{}
\begin{document}
\maketitle
\begin{abstract}
We present a comparative study of the dynamical behaviour of topological 
systems of recent interest, namely the non-Abelian Chern-Simons Higgs system 
and the Yang-Mills Chern-Simons  Higgs system. By reducing the full field 
theories to temporal differential systems using the assumption of spatially 
homogeneous fields , we study the Lyapunov exponents for two types of initial 
conditions. We also examine in minute detail the behaviour 
of the Lyapunov spectra as a function of the various coupling parameters in the 
system.
We compare and contrast our results with those for Abelian Higgs, Yang-Mills 
Higgs and Yang-Mills Chern-Simons systems which have been discussed by other 
authors recently. The role of the various terms in the Hamiltonians for such 
systems in determining the order-disorder transitions is emphasized and shown 
to be counter-intuitive in the Yang-Mills Chern-Simons Higgs systems.

\end{abstract}
\pagebreak 

\oddsidemargin -.1cm
\baselineskip 24pt
\section{Introduction}
  In recent times, the theory of dynamical systems has provided much insight 
into the origin of chaos in classical systems which were traditionally thought 
of as being completely deterministic. However, much of the progress has been 
mainly in the context of discrete mapping and those differential dynamical 
systems having low dimensional phase spaces.                                    
  
In the context of mathematical physics , there exist differential dynamical 
systems described by a large number of variables and therefore having phase
spaces of rather large dimensions. Examples of such systems are the Yang-Mills 
system (YM), the Chern-Simons system (CS) and their various enlargements such 
as the Yang-Mills Higgs (YMH) , Yang-Mills Chern-Simons Higgs (YMCSH) and 
Chern-Simons Higgs (CSH) systems. These systems are treated as dynamical 
systems after being derived from the original highly nonlinear partial 
differential equations (PDE'S) through the assumption of spatial homogeneity 
which reduces the dependence of the dynamical variables on the three or 
four-dimensional space-time coordinates to a dependence purely on time. 
The systems then become temporal differential dynamical systems.

Savvidy et.al \cite{sav} were the the first to demonstrate the chaotic nature 
of gaugetheories  by treating them as differential dynamical systems. In fact, 
a detailed investigationhas been conducted in this context \cite{savnuc} to 
classify the dynamical version of the pure YM field theory in terms of its 
ergodicity properties.

While the condition for ergodic behaviour of a system is that the generic phase
trajectory visits all regions of phase space given a sufficiently long time,
and all phase-averages can be replaced by time-averages, the YM system
exhibits stronger stochasticity properties in phase space. Investigations 
revealthat it is definitely a mixing system, i.e., no time-averaging is 
required to achieve `equilibrium'. In contrast to systems which are simply 
ergodic its spectrum is continuous. Indeed, the studies seem to indicate that 
the YM dynamical system is a Kolmogorov K-system, which exhibits stronger 
mixing properties than those mentioned above. A connected neighbourhood
of phase trajectories in this case, exhibits a positive average 
rate of exponential divergence or net positive Lyapunov exponent. Equivalently, 
by a remarkable theorem \cite{Ksystem} a K system has positive Kolmogorov-Sinai 
(KS) entropy which is a measure of the degree of chaos, analogous to entropy 
as a measure of disorder in stastistical mechanics. In fact, while the KS 
entropy itself is not straightforward to measure, the fact that it is a sum of 
the positive Lyapunov exponents implies that it is an important concept in the
classification of chaotic systems.

Another aspect to the study of gauge theories as dynamical systems is the
attempt to understand the ground (vacuum) state structure 
 of quantum chromodynamics (QCD), and also its behaviour in extreme 
 environments (such as high temperature). In this context, systems such as YMH, 
 YMCS and YMCSH have been studied \cite{bam}.

The addition of the CS term to various Abelian
and non-Abelian gauge theories leads to novel features in general, as it is
a topological term. Even in the complete PDE's, while a
study of the YM case reveals interesting results in connection with the
geometry and topology of four-dimensional manifolds, the related CS PDE's
have yielded information about three-dimensional manifolds. The symplectic
structure of CS theories differs in important ways from that of the Maxwell
or YM gauge theories. In perturbative gauge theories the pure CS theories
exhibit features that are absent in the Abelian gauge theory with both the 
Maxwell and CS terms. Delicate aspects relating to the infrared and ultraviolet 
behaviour and the regularization dependence in such perturbative theories
\cite{Deser}, the natural connection of quantized three-dimensional CS gauge 
theories with two-dimensional conformal field theories \cite{Bos}, and the 
effect of the YM term which acts as a singular perturbation when added to the 
SU(2) CS theory \cite{Nash}, have all been extensively explored in the 
literature.

Another aspect of interest in CS theories relates to its quantum mechanics.
Non-perturbative quantum mechanical anomalies in these theories
\cite{Dun}, infinite dimensional symmetry groups that arise 
in CS quantum mechanics \cite{Flo}, self-dual CS theories and extended 
supersymmetry \cite{Lee} are a few areas in which distinct signatures of CS 
theories in sharp contrast to those of other gauge theories have been reported.
 
In this paper we report on yet another aspect of CS theories and contrast
them with other gauge theories. This pertains to the chaotic nature of gauge
theories mentioned earlier. For our purpose, the equations of motion are made 
to evolve only temporally, by suppressing the spatial dependence. While the 
resulting equations are like the continuum analogues of discrete maps (the 
latter being an area where extensive work has been done on their chaotic 
behaviour), they are also the dynamical version of the
full gauge theory and hence represent one sector of the corresponding field
theory. In the literature this is used as a convenient reduction to examine
the integrability properties of the theory. This is because if this
sector is proven to be non-integrable the corresponding field theory also
will be chaotic \cite{Nik}.
 
   In our recent papers, we have shown that the Abelian CSH system without a 
kinetic term is integrable, while the inclusion of a kinetic term, making it 
into a Maxwell Chern-Simons Higgs (MCSH) system (or Yang-Mills with a U(1) 
symmetry group) yielded a non-integrable system which admitted chaos 
\cite{our1}. The systems were also examined for the Painlev\'e property. A 
numerical study of the Lyapunov exponents and phase space trajectories were 
carried out to show the existence of chaos in these systems. In \cite{our2}, 
the analysis was extended to the non-Abelian CSH and YMCSH system with an 
SU(2) symmetry group and both were found to be chaotic.
   
In all these studies, one aspect that deserves more attention is the 
possibility of observing the existence of a phase transition, i.e., is there 
a sharp order to chaos transition in the parameter space of these theories?
Of course in these systems energy can also be used as a parameter since it is 
a conserved quantity. Indeed in ref.\cite{gian} , it has been claimed that an 
order-chaos transition is seen within a narrow range of energies.
More recently, Kawabe \cite{kawab} has argued that the Abelian Higgs system 
(YMH with a U(1) symmetry group) also shows an order-chaos transition for 
certain ratios of the two parameters in the theory. Our paper examines this 
aspect of chaos in YM systems by studying the non-Abelian CSH (NACSH) and the 
YMCSH systems with an SU(2) symmetry group. A comparative study is done to see 
the role of the kinetic term, the Higgs term and the CS term in the transition.

Some interesting features regarding the details of the phase transition from 
order to chaos in the dynamical analogues of both Abelian and non-Abelian 
gauge theories have been reported in the literature. In the context of Abelian 
Higgs theories Kawabe has reported transition from order to chaos
within a certain range of the Higgs coupling constant and energy \cite{kawab}. 
The onset of chaos is remarkably different qualitatively from the corresponding 
transition in the YMCS system where Giansanti and Simic \cite{gian} have 
reported the existence of an interesting `fractal' structure in the phase 
transition region. Much earlier Savvidy et. al \cite{mat} have observed that 
the role of the Higgs is to order the Yang-Mills system and later extensive 
work on the YMH systems was conducted in ref \cite{ohta} . The picture that 
emerges therefore, is that the Higgs and the Chern-Simons term have distinct 
and different roles to play in the transition.
It is therefore of importance to examine the effect of both terms on the Yang 
-Mills field in the YMCSH theory.
As a primer to this, the competing effect of the `oscillatory' behaviour of 
the CS term and the `stabilizing' role
of the Higgs term in the CSH system is investigated in this paper. Later,  we 
contrast this with the corresponding 
results in the YMCSH system. It is thus obvious that a K system like the 
Yang-Mills theory when coupled to CS and Higgs must
show a rich and instructive behaviour in the understanding of regularity vs. 
chaos in Hamiltonian systems.
 A second aspect which emerges is related to an interesting question - will the 
Higgs stabilize
 any gauge invariant term involving only the vector potentials, independent of 
whether 
 it is of the YM type or the CS type or is it necessary that the gauge
field is 
only YM in nature? We attempt to
 answer these questions in this paper.
\section{The Dynamical Systems}
In this section we set up the two systems we shall be examining. Let us consider
first the non-Abelian pure Chern-Simons Higgs system (i.e., without the kinetic 
term).
From our earlier paper  \cite{our2},
the Lagrangian for the non-Abelian (SU(2))
 CSH (NACSH) system in 2+1 dimensions
in Minkowski space is given by:
\begin{equation}
L = \frac{m}{2}\epsilon^{\mu\nu\lambda}[F^{a}_{\mu\nu}A^{a}_{\alpha} -
\frac{g}{3}f_{abc}A^{a}_{\mu}A^{b}_{\nu}A^{c}_{\alpha}]+D_{\mu}
\phi^{\dagger}_{a}D^{\mu}\phi_{a}-V(\phi)
\end{equation}
where
\begin{equation}
F^{a}_{\mu\nu}= \partial_{\mu}A_{\nu}^{a}-\partial_{\nu}A_{\mu}^{a}
+gf_{abc}A^{b}_{\mu}A^{c}_{\nu},
\end{equation}
$f_{abc}$ are the structure constants of the SU(2) Lie algebra
and
\begin{equation}
D_{\mu}\phi_{a}=(\partial_{\mu}-igT^{l}A^{l}_{\mu})\phi_{a}.
\end{equation}
Here, $T_{a} $ are the generators of the SU(2) algebra, so that  
$ tr[T_{a}T_{b}]=\delta_{ab} $.
The equations of motion become:
\begin{eqnarray}
m\epsilon^{\nu\alpha\beta}F^{a}_{\alpha\beta}=ig[D^{\nu}
\phi^{\dagger}T_{a}\phi - \phi^{\dagger}T_{a}D^{\nu}\phi]  \\
D_{\mu}D^{\mu}(\phi)=-\frac{1}{2}\frac{\partial V}{\partial\phi}
\end{eqnarray}
Then, considering the real triplet representation for
the Higgs field  and the spatially homogeneous solutions
$ \partial_{i}A_{j}^{a}=\partial_{i}\phi=0$ i,j=1,2
and the gauge choice $A_{0}^{a}=0$ we get for the $\nu=0$ component of 
eqn.[4]
\begin{equation}
m \vec{A}_{1}\times\vec{A}_{2}=-\vec{\phi}\times\vec{\dot{\phi}}
\end{equation}
which is just the Gauss' law constraint.
The remaining equations of motion for the vector field are:
\begin{eqnarray}
\dot{\vec{A_{1}}}=\frac{g^{2}}{m}(\vec{A_{2}}\vec{\phi^{2}}-\vec{\phi}
\vec{A_{2}}\cdot \vec{\phi})\\
\dot{\vec{A_{2}}}=-\frac{g^{2}}{m}(\vec{A_{1}}\vec{\phi^{2}}-\vec{\phi}
\vec{A_{1}}\cdot \vec{\phi})
\end{eqnarray}
The equation of motion for the Higgs field is:
\begin{equation}
\ddot{\vec{\phi}} =-g^{2}[(\vec{A_{1}^{2}}+\vec{A_{2}^{2}})\vec{\phi}-
(\vec{A_{1}}
\cdot\vec{\phi}\vec{A_{1}}+\vec{A_{2}}\cdot\vec{\phi}\vec{A_{2}}]-\frac{1}{2}
\frac{\partial V}{\partial \vec{\phi}}.
\end{equation}
From the equations of motion for the vector fields it is easily seen that
$\vec{A_{1}^{2}}+\vec{A_{2}^{2}}$ is a constant of the motion. Throughout
this paper, we work with the potential
\begin{equation}
V(\phi) = \frac{\lambda}{4}(\vec{\phi}^{2}-v^{2})^{2}.
\end{equation}

The NACSH system described by the above equations of motion has three
parameters. For comparison with the work of Kawabe \cite{kawab}, we shall 
scale the variables such that we are left with only one parameter.
The following scaling of variables,
\[ \vec{A_{1}}\longrightarrow g\vec{A_{1}} \]
\[ \vec{A_{2}}\longrightarrow g \vec{A_{2}} \]
\[ \vec{\phi}\longrightarrow \frac{g}{\sqrt{m}}\vec{\phi} \]
\[ v\longrightarrow \frac{g}{\sqrt{m}}v \]
reduces the equations of motion to:
\begin{eqnarray}
\vec{\dot{A_{1}}}&=&[\vec{A_{2}}\vec{\phi^{2}}-\vec{\phi}(\vec{A_{2}}\cdot
\vec{\phi})]\\
\vec{\dot{A_{2}}}&=&-[\vec{A_{1}}\vec{\phi^{2}}-\vec{\phi}(\vec{A_{1}}\cdot
\vec{\phi})]\\     
\vec{\ddot{\phi}}&=&-[(\vec{A_{1}^{2}}+\vec{A_{2}^{2}})\vec{\phi}-
(\vec{A_{1}}\cdot\vec{\phi}\vec{A_{1}}+\vec{A_{2}}\cdot\vec{\phi}\vec{A_{2}})]-
\frac{\kappa}{2}\vec {\phi}(\vec{\phi}^{2}-v^{2})                
\end{eqnarray}

with $\kappa=\frac{\lambda m}{g^{2}}$ . In the rest of the paper,
we shall set the scaled v to be one without loss of generality.
The scaled Lagrangian is given by
\begin{equation}
L =(\vec{\dot{A_{1}}}\cdot \vec{A_{2}}-\vec{\dot{A_{2}}}\cdot\vec{A_{1}})
+\vec{\phi}^{2}
-[(\vec{A_{1}^{2}}+\vec{A_{2}^{2}})\vec{\phi}^{2}-(\vec{A_{1}}\cdot
\vec{\phi})^{2}-(\vec{A_{2}}\cdot\vec{\phi})^{2}]-\frac{\kappa}{4}
(\vec{\phi}^{2}-1)^{2}  
\end{equation}
The corresponding energy function is easily seen to be 
\begin{equation}
E =\vec{\dot{\phi}}^{2}+[(\vec{A_{1}^{2}}+\vec{A_{2}^{2}})\vec{\phi}^{2}
-(\vec{A_{1}}\cdot\vec{\phi})^{2}-(\vec{A_{2}}\cdot\vec{\phi})^{2}]
+\frac{\kappa}{4}(\vec{\phi} ^{2}-1)^{2}  .
\end{equation}
These are the equations governing the NACSH dynamical system. Since we wish 
to compare the results
for the NACSH system with those of the YMCSH system we proceed to set up the 
YMCSH dynamical equations.
The Lagrangian is given by
 \begin{eqnarray}
L&=&-\frac{1}{4}F_{\mu\nu}^{a}F^{\mu\nu a}+\frac{m}{2}\epsilon^{\mu\nu\alpha}
[F_{\mu\nu}^{a}A^{a}_{\alpha}-\frac{g}{3}f_{abc}A^{a}_{\mu}A_{\nu}^{b}
A^{c}_{\alpha}]\nonumber \\
&&+D_{\mu}\phi^{\dagger}_{a}D^{\mu}\phi_{a}-V(\phi)
\end{eqnarray}
The equations of motion are
\begin{eqnarray}
D_{\mu}F^{\mu\nu a}+m\epsilon^{\nu\alpha\beta}F^{a}_{\alpha\beta}=ig[D^{\nu}
\phi^{\dagger}T_{a}\phi-\phi^{\dagger}T_{a}D^{\nu}\phi]\\
D_{\mu}D^{\mu}\phi^{a}=-\frac{\partial V}{\partial \phi^{*}_{a}}=-\frac{1}{2}
\frac{\partial V}{\partial \phi_{a}}
\end{eqnarray}
The $\nu=0$ component gives the Gauss' law constraint which in this case is
\begin{equation}
\frac{1}{2}(\vec{A_{1}}\times{\dot{\vec{A_{1}}}}+\vec{A_{2}}\times\dot
{\vec{A_{2}}}+2m\vec{A_{1}}\times\vec{A_{2}})=-2\vec{\phi}\times
{\vec{\dot{\phi}}}.
\end{equation}
Once again, we choose the gauge $A_{0}=0$ and consider the spatially homogeneous
 case ; then the equations of motion for the gauge fields become:
\begin{eqnarray}
\ddot{\vec{A_{1}}}+2m\dot{\vec{A_{2}}}+2{g^{2}}(\vec{A_{1}}\vec{\phi^{2}}-
\vec{\phi}\vec{A_{1}}\cdot \vec{\phi})\nonumber \\
+g^{2}(\vec{A_{1}}\vec{A_{2}}\cdot\vec{A_{2}}-\vec{A_{2}}\vec{A_{1}}\cdot
\vec{A_{2}})=0\\
\ddot{\vec{A_{2}}}-2m\dot{\vec{A_{1}}}+2{g^{2}}(\vec{A_{2}}\vec{\phi^{2}}-
\vec{\phi}\vec{A_{2}}\cdot \vec{\phi})\nonumber \\
+g^{2}(\vec{A_{2}}\vec{A_{1}}\cdot\vec{A_{1}}-\vec{A_{1}}\vec{A_{1}}\cdot
\vec{A_{2}})=0.
\end{eqnarray}
From these equations it is easy to see that
\[\vec{A_{2}}\cdot\dot{\vec{A_{1}}}-\vec{A_{1}}\cdot\dot{\vec{A_{2}}}+
m(\vec{A_{1}^{2}}+\vec{A_{2}^{2}})\]
is a constant of the motion.
 The equation of motion for the three component Higgs field becomes
\begin{equation}
\ddot{\vec{\phi}} =-g^{2}[(\vec{A_{1}^{2}}+\vec{A_{2}^{2}})\vec{\phi}-
(\vec{A_{1}}
\cdot\vec{\phi}\vec{A_{1}}+\vec{A_{2}}\cdot\vec{\phi}\vec{A_{2}})]-
\frac{1}{2}\frac{\partial V}{\partial \vec{\phi}}.
\end{equation}
In this case we have three second order differential equations for each vector 
field.
The Lagrangian which leads to these equations of motion is:
\begin{eqnarray}
L=\frac{1}{2}(\dot{\vec{A_{1}^{2}}}+\dot{\vec{A_{2}^{2}}})
+ m(\dot{\vec{A_{1}}}\cdot \vec{A_{2}}-\dot{\vec{A_{2}}}\cdot\vec{A_{1}})
+\dot{\vec{\phi}}^{2}\nonumber \\
-g^{2}[\frac{1}{2}(\vec{A_{1}^{2}}\vec{A_{2}^{2}}-(\vec{A_{1}}\cdot
\vec{A_{2}})^{2})+(\vec{A_{1}^{2}}+\vec{A_{2}^{2}})\vec{\phi^{2}}-(\vec{A_{1}}
\cdot\vec{\phi})^{2} -(\vec{A_{2}}\cdot\vec{\phi})^{2}]-V(\phi)
\end{eqnarray}
Using the same rescaling as for NACSH we have the equations of motion
\begin{eqnarray}
\frac{1}{m}\ddot{\vec{A_{1}}}+2\dot{\vec{A_{2}}}+2(\vec{A_{1}}\vec{\phi^{2}}
-\vec{\phi}\vec{A_{1}}\cdot \vec{\phi})\nonumber \\
+\frac{1}{m}(\vec{A_{1}}\vec{A_{2}}\cdot\vec{A_{2}}-\vec{A_{2}}\vec{A_{1}}
\cdot\vec{A_{2}})=0\\
\frac{1}{m}\ddot{\vec{A_{2}}}-2\dot{\vec{A_{1}}}+2(\vec{A_{2}}\vec{\phi^{2}}
-\vec{\phi}\vec{A_{2}}\cdot \vec{\phi})\nonumber \\
+\frac{1}{m}(\vec{A_{2}}\vec{A_{1}}\cdot\vec{A_{1}}-\vec{A_{1}}\vec{A_{1}}
\cdot\vec{A_{2}})=0
\end{eqnarray}
and
\begin{equation}
\ddot{\vec{\phi}} =-[(\vec{A_{1}^{2}}+\vec{A_{2}^{2}})\vec{\phi}-(\vec{A_{1}}
\cdot\vec{\phi}\vec{A_{1}}+\vec{A_{2}}\cdot\vec{\phi}\vec{A_{2}}]-
\frac{\kappa}{2}\vec{\phi}(\vec{\phi}^{2 }-1).
\end{equation}
It is interesting to note that while in the NACSH system the Yang-Mills
parameter
 g, the Higgs parameter $\lambda$ and the Chern-Simons parameter m could all be
 combined
into the parameter $\kappa$, this is not possible for the YMCSH system where we
 are left with both $\kappa$ and m appearing explicitly.
The scaled Lagrangian in this case is given by
\begin{eqnarray}
L=\frac{1}{2m}(\dot{\vec{A_{1}^{2}}}+\dot{\vec{A_{2}^{2}}})+ 
(\dot{\vec{A_{1}}}\cdot \vec{A_{2}}-\dot{\vec{A_{2}}}\cdot\vec{A_{1}})
+\dot{\vec{\phi}}^{2}\nonumber \\
-\frac{1}{2m}[\vec{A_{1}^{2}}\vec{A_{2}^{2}}-(\vec{A_{1}}\cdot
\vec{A_{2}})^{2}]+[(\vec{A_{1}^{2}}+\vec{A_{2}^{2}})\vec{\phi^{2}}-(\vec{A_{1}}
\cdot\vec{\phi})^{2}-(\vec{A_{2}}\cdot\vec{\phi})^{2}]-\frac{\kappa}{4}
(\vec{\phi}^{2}-1)^{2}
\end{eqnarray}
with an energy function 
\begin{equation}
E=\frac{1}{2m}(\dot{\vec{A_{1}^{2}}}+\dot{\vec{A_{2}^{2}}})+ 
\vec{\dot{\phi}}^{2}
+\frac{1}{2m}[\vec{A_{1}^{2}}\vec{A_{2}^{2}}-(\vec{A_{1}}\cdot\vec{A_{2}})^{2}]
+[(\vec{A_{1}^{2}}+\vec{A_{2}^{2}})\vec{\phi^{2}}-(\vec{A_{1}}\cdot
\vec{\phi})^{2}-(\vec{A_{2}}\cdot\vec{\phi})^{2}] +\frac{\kappa}{4}
(\vec{\phi}^{2}-1)^{2}.
\end{equation}
This completes the description of the dynamical systems which we shall be
studying.
In the next section, we describe the numerical analysis that we have carried 
out.
\section{Numerical Analysis}
It is clear from the above dynamical equations that the phase space of these 
systems are large. One of the traditional ways for examining
phase spaces to determine chaotic vs. regular behaviour has been to use the 
Poincar\'e sections where one examines the points mapped 
out on a plane surface in phase space as the trajectory crosses it.
While this is straightforward  for dynamical systems whose phase space 
dimensionality does
not exceed four, it is difficult to interpret it in the systems we are dealing 
with.
  
Instead, we examine the variation of the maximal Lyapunov exponent as the two 
NACSH parameters  
energy and $\kappa$ are varied. This clearly shows us regions of regular 
behaviour (where the exponent goes to zero)
and regions of chaotic behaviour (where the exponent is positive). These
calculations were carried out for a wide range of initial
conditions. The initial conditions that were chosen were in turn dictated by the
 dynamical systems themselves. Being derived from the
equations of motion the field variables are required to satisfy the Gauss' law 
constraint.
Since this constraint must be preserved during the time evolution of the
dynamical system, it is sufficient to ensure their validity via the initial 
conditions.

For the NACSH system the following forms were chosen as the initial conditions:
\begin{equation}
\vec{A_{1}}=\left( \begin{array}{c} x \\ 0\\0  \end{array}\right) \;\;\;\;
\vec{A_{2}}=\left( \begin{array}{c} 0\\x\\0  \end{array}\right) \;\;\;\;
\vec{\phi}=\left( \begin{array}{c} -x\\x\\0  \end{array} \right) \; \;\;\;
\dot{\vec{\phi}}=\frac{1}{2}\left(\begin{array}{c}x\\ x \\ 0 \end{array}\right)  .
\end{equation}
For YMCSH this is supplemented with
\begin{eqnarray}
\dot{\vec{A_{1}}}=\left( \begin{array}{c} 0 \\0\\0  \end{array}\right)\;\;\;\;
\dot{\vec{A_{2}}}=\left( \begin{array}{c} 0\\0\\0  \end{array} \right)  . 
\end{eqnarray}
x was then varied to obtain a range of initial conditions for suitable
energies of interest.
In figs. 1--4 we show the behaviour of the maximal Lyapunov exponent as a 
function
of energy for $\kappa=0,.5,1,5$. For $\kappa=0$ (i.e., absence of the Higgs 
potential)
we see that the system is mostly {\bf chaotic} with a window of regularity for 
$7 \le E \le 9$.
Increasing the value of $\kappa$ , we find a transitional region of regular to 
chaotic behaviour
at small energies, in contrast to mostly oscillatory behaviour at higher 
energies. For $\kappa=1$, more transitions from order to chaos appear
at  larger energies. Much more dramatic behaviour is seen for large 
$\kappa$ ($\kappa=5$), where oscillatory
behaviour manifests itself for $E\ge 3$.
Thus we see that the effect of the topological term (large m) is to produce  
a regular oscillatory behaviour in the dynamics of the system.
  
To illustrate the regular and chaotic behaviour on the trajectories of this 
system
we exhibit in figs. 5 and 6, phase plots corresponding to the energies at 
which
the Lyapunov exponent shows regular and chaotic behaviour. These correspond
to 
$\kappa=1$ and energies 10.28 and 20.607.
  
Giansanti and Simic \cite{gian} report a fractal-like structure of 
order-chaos transitions in YMCS systems. In our particular case, when the Yang-
Mills field is absent
and the Higgs field is present we see no such fractal behaviour in the region
of phase space that we have examined. Therefore, this 
suggests that the quartic
coupling that arises from the inclusion of the kinetic YM term may be
responsible
for the observed fractal structure. This rich phase space structure of the CSH, 
YMCS and YMCSH systems clearly
needs further exploration. The set of trajectories examined by Giansanti and 
Simic do not 
correspond to the ansatz that we have chosen and therefore we have explored 
different regions
of phase space. Thus our results are complementary to those obtained by 
Giansanti and Simic.
  
We have examined the NACSH system for another ansatz which we may call the two-v
 variable ansatz:
\begin{eqnarray}
\vec{A_{1}}= \left( \begin{array}{c} x \\0\\0  \end{array}\right) \;\;\;\;
\vec{A_{2}}=\left( \begin{array}{c} 0\\y\\0  \end{array}\right) \;\;\;\;
\vec{\phi}=\left( \begin{array}{c} -x\\y\\0  \end{array}\right) \; \;\;\;
\vec{\dot{\phi}}=\frac{1}{2} \left( \begin{array}{c} x\\ y \\ 0 \end{array} 
\right)  .
\end{eqnarray}
For YMCSH this is supplemented with
\begin{eqnarray}
\vec{\dot{A_{1}}}= \left( \begin{array}{c} 0 \\0\\0  \end{array}\right) 
\;\;\;\;
\vec{\dot{A_{2}}}=\left( \begin{array}{c} 0\\0\\0  \end{array} \right)  .
\end{eqnarray}
This allows us to compare our results with that of Kawabe for the Abelian Higgs 
theory \cite{kawab},
where motion is completely bounded for $Q=\frac{4Eg^{2}}{\lambda}\le 1$, while 
it is unbounded for
$Q>1$. In contrast, for the CSH system that we are studying ,  for all values of 
$Q=4E\nolinebreak\slash\kappa \hphantom{...}$   and  $\hphantom{...}\kappa=
\frac{\lambda}{mg^{2}}$, the motion
is bounded. Indeed we find that for large E and large $\kappa$ the contour 
levels
of the function 
$W=[(\vec{A_{1}}^{2}+\vec{A_{2}}^{2})\vec{\phi}^{2}-(\vec{A_{1}}
\cdot\vec{\phi})^{2}-(\vec{A_{2}}\cdot\vec{\phi})^{2}]+\frac{\kappa}{4}
(\vec{\phi}^{2}-1)^{2}$ show extremely restricted domains for the dynamics.
Fig. 7 represents the potential contours for $\kappa=10$ for energies in the 
range $\{0.25,2\}$ while fig. 8 gives the contours for
the energy range $\{2,10\}.$
There is a dramatic change in the available phase space for this particular two-
variable ansatz.
This clearly suggests that the non-Abelian nature of the CS term is already 
contributing to a significant
change in the dynamical structure.

Evidence for the transition from regularity to chaos is seen in fig. 9, where we
show the fraction of the phase space that is regular. This is obtained by 
calculating
the maximal Lyapunov exponent for $\kappa=1$ for various initial conditions for 
energies ranging from 1 to 10. The cutoff on the exponent
for regularity was taken to be 0.01. A simple count on the exponents falling 
below this value out of a hundred
initial conditions for each energy was carried out. The figure clearly reveals 
that for small $\kappa$ , the NACSH system becomes chaotic more or less 
monotonically as the energy is increased. However for 
large $\kappa$ there is no such simple behaviour . A fuller discussion of this 
is given in the next section.

This transition between regular and chaotic behaviour is also seen in the phase 
space trajectories shown in figs. 10 and 11. In fig. 10 we plot 
$\phi_{2}$ vs. $\phi_{1}$
for $\kappa=1$, energy=10 and an initial condition where the maximal Lyapunov 
exponent is almost zero. In fig. 11 we show the phase plot for the same parameter 
values for an initial condition which gives a large maximal 
Lyapunov exponent.
  
We now proceed to the numerical investigation of the YMCSH systems. Fig. 12 
shows the variation of the maximal Lyapunov exponent as a function of energy
for $\kappa=1,10$ and energy=10, with m=1. The graphs clearly show
that for large $\kappa$ (where either the Higgs coupling $\lambda$ is large or 
the YM coupling g is small) the system exhibits more regularity for low energies.

This is to be contrasted with earlier work \cite{sav},\cite{kawab},where it was 
reported that in a YMH system as $\lambda$ increases the system becomes more 
regular regardless of the energy regime being investigated.
Here, we see that for the YMCSH system such transitions to regularity are seen 
for small energies as $\kappa$ increases while the Lyapunov exponent increases 
almost linearly with energy for the large energy regime.
While fig.12 was obtained for the one-variable ansatz, fig.13 shows the maximal 
Lyapunov exponent as a function of the initial variable x  for the two-variable 
ansatz, for fixed values of energy and $\kappa$.
Once again we see regions of regularity for small energies and chaotic behaviour
 for large energies irrespective of the value of $\kappa$. Evidence for 
regularity and chaos for the two-variable ansatz is given in figs.13 and 14.
In fig.14 we see that for $\kappa=5$ and energy=1, the phase space trajectory is
 highly regular and quasi-periodic. Fig.15 shows a region for the same parameter
 values, where the phase space is chaotic.
  
All the calculations for the NACSH system were carried out using a 
straightforward Runge-Kutta fourth order routine with care being taken to 
preserve the constants of motion to an accuracy of one part in $10^{5}$.
However, the calculations for the YMCSH system required an adaptive step-size 
Runge-Kutta routine to ensure energy conservation to the same degree of 
accuracy as in the NACSH system.
\section{Results and Discussion}
From the numerical studies undertaken by us, certain very interesting
features emerge, not only regarding the richness of the phase space
corresponding to gauge theories, but also with respect to some 
`counter-intuitive' phenomena that occur in these systems.

One feature that seems to be common to both the YMH \cite{mat} and CSH systems, 
from an
investigation of the potential contours corresponding to various initial
conditions, is the boundedness of phase space.
This is in contrast to the situation that prevails in the Abelian Higgs 
system \cite{kawab}. The boundedness observed is better understood when we realize 
that the YMCS dynamical system can be used to describe particle motion in a 
magnetic field \cite{gian}, a physical situation which allows only for bounded 
motion.

Apart from the crucial role played by the ansatz chosen by us, in this matter, 
the non-Abelian
nature of the gauge term also plays an important part. In this context, we 
realize that the boundedness observed
is quite independent of the details of the non-Abelian coupling, i.e., whether 
it is of the YM type or of the trilinear CS type.

We recall that the YM system is a K-system \cite{savnuc} and that the CS term 
creates more regular windows when added
to the YM system \cite{gian}. This accounts to a large measure, for the fact 
that for some $\kappa$ and m values,
the maximal Lyapunov exponents in the YMCSH case are larger than the 
corresponding ones for the CSH system. 

However, a striking feature that emerges in our studies that is 
counter-intuitive is that, in general, it is not true that
increase in $\kappa$ `regularizes' the gauge term at all energies. Recall that 
our calculations are made in terms of the scaled parameter $\kappa$.
An increase in $\kappa$ could either be due to an increase in the Higgs coupling
$\lambda$ or a decrease in the gauge coupling $g$, for a fixed m.
Whereas in the former case the regularity is expected to increase , in the 
latter case, it is not established that for all small
non-zero g more regular islands appear. In fact, as borne out by figs. 12,13 
and 14, {\bf an increase in $\kappa$ produces more regularity only for small 
values of energy}.
This can be understood better if we realize that it is not just the value of 
$\kappa$ that determines the appearance of regular islands, but
 also the available phase space as well. This is clear from the energy contours 
shown in figs.7 and 8. With more accessible regions
in phase space, it is but natural that more randomness sets in and the KAM tori 
structure gets broken.

Another aspect is that as the energy increases, the maximal Lyapunov exponent 
increases in magnitude in the YMCSH system. This shows that the YM terms takes 
over for large energies and the CS term produces the `oscillatory effect'. The 
effect of the CS term is reminiscent of the `fractal' structure observed in 
YMCS systems,
where in various energy windows, order-chaos-order transitions are observed 
\cite{gian}.

An important feature that we see is that while in the pure CSH system large 
windows of regularity exist ( as seen for the range of energies shown in 
figs.1-4), in  YMCSH this does not happen.
This perhaps can be traced back to the interplay between the oscillatory effect 
of the CS term, the regularizing effect of the Higgs term and the completely 
chaotic nature of the YM term.
Further , with increase in $\kappa$ , transitions from regularity to chaos take 
place in the pure CSH case at relatively higher energies when compared with the 
low energy values 
corresponding to order-chaos transitions in YMCSH.

The final picture which emerges bears out the fact that in a complex 
dynamical system with a large phase space (in contrast to the wide class 
of Hamiltonian systems with two degrees of freedom) curious interplay between 
different coupling constants and the rich structure of phase space itself can 
lead to novel results- some of them quite counter-intuitive and surprising. A 
detailed examination of dynamical systems which emerge from field theoretic 
systems are thus vital for an understanding of Hamiltonian systems with a large 
number of degrees of freedom. In turn it is also an important primer in the 
understanding of non-Abelian field theories themselves.
\section{Acknowledgments}
C.M  wishes to thank the Indian National Science Academy(INSA) for the award of 
an INSA visiting fellowship. He would also like to thank the University Grants 
Commission (UGC,India) for partial support 
under its research scientists scheme. M.S.S would like to thank the UGC for 
partial support under its visiting associate scheme. J.S thanks the Department 
of Atomic Energy (DAE, India) for a fellowship.
B.B acknowledges the UGC for partial support under its scientist scheme.
S.L is very grateful for hospitality and use of the PACE+ computer during her 
stay in ANURAG, Hyderabad. She especially acknowledges help given by K.Mahesh 
and S.Seshadri.
We are all grateful to Dr. R.Nagarajan for his assistance, and to the
referees for their useful comments.
\pagebreak

\pagebreak

\section{Figure Captions}
Fig.1 Variation of the maximal Lyapunov exponent with energy for NACSH for 
$\kappa=0$.\\
Fig.2 Variation of the maximal Lyapunov exponent with energy for NACSH for 
$\kappa=0.5$.\\
Fig.3 Variation of the maximal Lyapunov exponent with energy for NACSH for 
$\kappa=1.0$.\\
Fig.4 Variation of the maximal Lyapunov exponent with energy for NACSH for 
$\kappa=5.0$.\\
Fig.5 Phase plot of $\phi_{2}  vs. \phi_{1}$ showing regular behaviour for NACSH 
for the one-variable ansatz for $\kappa=1$, E=10.2088 and x=1.381.\\
Fig.6 Phase plot of $\phi_{2}  vs. \phi_{1}$ showing chaotic behaviour for NACSH 
for the one-variable ansatz for $\kappa=1$, E=20.6071 and x=1.64.\\
Fig.7 Contour plot for the potential corresponding to NACSH for $\kappa=10$ and 
energy range $\{.25,2\}$\\
Fig.8 Contour plot for the potential corresponding to NACSH for $\kappa=10$ and 
energy range $\{2,10\}$\\
Fig.9 Phase space fraction of regularity vs. energy for $\kappa=1$ for NACSH\\
Fig.10 Phase plot of $\phi_{2}  vs. \phi_{1}$ showing regular behaviour for NACSH
for the two-variable ansatz for $\kappa=1$, E=10,  x=.05 and y=2.598405\\
Fig.11 Phase plot of $\phi_{2}  vs. \phi_{1}$ showing chaotic behaviour for NACSH
for the two-variable ansatz for $\kappa=1$, E=10, x=.9 and y=1.908953.\\
Fig.12 Comparison of maximal Lyapunov exponent vs. energy for YMCSH for the one-
variable ansatz for $\kappa=1$, m=1
($\ast$ represents the curve) with $\kappa=10$, m=1 (+ represents the curve).\\
Fig.13 Comparison of maximal Lyapunov exponent vs. x for YMCSH for the 
two-variable ansatz for $\kappa=5$, E=1,m=1
(o represents the curve) with $\kappa=10$,E=5, m=1 (+ represents the curve) and 
$\kappa=1$, E=10, m=1 ($\ast$ represents the curve)\\
Fig.14 Phase plot of $\phi_{2}  vs. \phi_{1}$ showing regular behaviour for YMCSH
 for the two-variable ansatz for $\kappa=5$, E=1,  x=.725 and y=0.742085.\\
Fig.15 Phase plot of $\phi_{2}  vs. \phi_{1}$ showing chaotic behaviour for YMCSH
 for the two-variable ansatz for $\kappa=5$, E=1,  x=.9125 and y=0.571046.\\
\end{document}